\documentstyle [12pt] {article}

\newcommand{\gtwid}{\mathrel{\raise.3ex\hbox{$>$\kern-.75em\lower1ex
\hbox{$\sim$}}}}
\newcommand{\ltwid}{\mathrel{\raise.3ex\hbox{$<$\kern-.75em\lower1ex
\hbox{$\sim$}}}}
\newcommand{\beq}{\begin{equation}}
\newcommand{\eeq}{\end{equation}}
\newcommand{\be}{\begin{equation}}
\newcommand{\ee}{\end{equation}}
\newcommand{\beqs}{\begin{eqnarray}}
\newcommand{\eeqs}{\end{eqnarray}}
\def\({\left (}
\def\){\right )}

\begin{document}

\begin{titlepage}

\begin{flushright}
\begin{tabular}{l}
ITP-SB-94-55    \\
December, 1994
\end{tabular}
\end{flushright}

\vspace{8mm}
\begin{center}
{\Large \bf Models of Fermion Mass Matrices Based on a Flavor- and
Generation-Dependent U(1) Gauge Symmetry}

\vspace{4mm}
\vspace{16mm}

Vidyut Jain \footnote{email: vid@max.physics.sunysb.edu}
and Robert Shrock\footnote{email: shrock@max.physics.sunysb.edu}

\vspace{6mm}
Institute for Theoretical Physics  \\
State University of New York       \\
Stony Brook, N. Y. 11794-3840  \\

\vspace{20mm}

{\bf Abstract}
\end{center}

We study models of fermion mass matrices based on a flavor- and
generation-dependent string-motivated U(1)$_A$ gauge symmetry and report
two new classes of solutions to the requisite consistency conditions.  In
particular, we propose that the fundamental reason underlying the striking
feature $m_b, \ m_\tau << m_t$ is that all of the elements of the down-quark
and charged lepton effective Yukawa matrices actually arise from
higher-dimension operators, suppressed by inverse powers of the Plank mass. An
explicit model embodying this idea is constructed.

\vspace{35mm}

 \end{titlepage}
\newpage
\setcounter{page}{1}
\pagestyle{plain}
\pagenumbering{arabic}
\renewcommand{\thefootnote}{\arabic{footnote}}
\setcounter{footnote}{0}

   The pattern of fermion masses and mixing remains one of the most important
mysteries in particle physics.  The successful standard model (SM) can
accomodate but not explain this pattern. A satisfactory understanding
would require that one have an experimentally confirmed theory explaining the
related electroweak symmetry breaking (EWSB), and one does not have this at
present.  Nevertheless, one may proceed by exploring plausible models.  The
fermion mass spectrum has several striking features: (i) within each charge
sector, the masses increase with generation by large factors:
$m_u << m_c << m_t$, $m_d << m_s << m_b$, and $m_e << m_\mu << m_\tau$; (ii) if
one assumes that all of these masses arise from conventional, dimension-4
Yukawa couplings, the associated Yukawa couplings for all of these fermions
except the top quark are all much smaller than a typical small coupling like
$e=\sqrt{4\pi\alpha} \simeq 0.3$, without explanation; (iii) related to (ii),
even if one restricts to the third generation, the masses are still quite
different: $m_\tau$ and $m_b$ are both $<< m_t$.  A related feature is that
(iv)  the Cabibbo-Kobayashi-Maskawa (CKM) quark mixing matrix is near to the
identity.  If one assumes certain simple forms for Yukawa matrices with
zeros in various entries, it is possible to explain property (iv) as a
consequence of (i), because the quark mixing angles are functions of (square
roots of) small mass ratios like $m_d/m_s$, $m_u/m_c$, etc. \cite{w,f}.
However, in such an approach, the fermion masses are used as inputs, and
properties (i), (ii), and (iii) are not explained.\footnote{Indeed, the model
of Ref. \cite{f} is excluded by property (iii), since in that model,
$|V_{cb}| \simeq |\sqrt{m_s/m_b} - e^{i\phi}\sqrt{m_c/m_t}|$ (masses normalized
at same scale $\mu$); because $m_t >> m_c$, the second
term is too small to significantly cancel the first term
$\sqrt{m_s/m_b} \simeq 0.17$ and fit the experimental value,
$|V_{cb}| = 0.038 \pm 0.005$.}  Indeed, although there has been recent progress
in explaining (i) and (iv)
via contributions of higher-dimension operators at a high
mass scale near to the scale of quantum gravity \cite{ir}, such efforts have
not addressed what, to us, seems an equally remarkable feature, viz., (iii).
The attractive idea of radiative electroweak symmetry
breaking in a supersymmetric generalization of the standard model
\cite{rad} depends on the existence of {\it at least} one quark
which has a mass comparable to the EWSB scale, but it cannot explain why this
was the top quark instead of the bottom quark (or, indeed why {\it both} $m_t$
and $m_b$ are not $\sim$ the EWSB scale), and hence it cannot explain
(iii) or the full extent of (ii).

   In this paper, we shall explore an appealing class of models of fermion
mass matrices which has the potential to explain all of the properties
(i)-(iv).  We shall
present a particular model which, we believe, is the first to offer a possible
fundamental explanation of property (iii).  The explanation is that in the
down-quark and charged lepton sectors, the masses of
not just the first two generations, but of
{\it all} generations arise from higher-dimension operators suppressed by
powers of a small mass ratio, $\epsilon \propto v/\bar M_P$, where $v$ is the
breaking scale of a
flavor- and generation-dependent, string-motivated U(1)$_A$ symmetry, and $\bar
M_P \equiv (8 \pi G_N)^{-1/2} = 2.44 \times 10^{18}$ GeV is the (reduced)
Planck mass.

   To set our work in context, we note that
since the original success of the standard model, there has been a growing
appreciation that its renormalizability and, in particular, the absence of
higher-dimension operators, may well be the consequence of a large logarithmic
interval in energy between the electroweak scale and a higher scale where there
is new physics (see, e.g., Ref. \cite{w92}).  Indeed, there are specific
reasons for expecting such operators at this high scale: the only known way to
stabilize the hierarchy $v_{_{EW}} << \bar M_P$ (where $v_{_{EW}}=246$ GeV
is the EWSB scale)  is via a supersymmetric
generalization of the SM.  In turn, global supersymmetry is naturally embedded
in supergravity, which one also finds as the low-energy limit of the main
candidate for quantum gravity, string theory.  But $d=4$ supergravity is
nonrenormalizable.  Indeed, explicit calculations of the
pointlike limit of string theories for energies $E << M_{str}$ (where
$M_{str} = 2(\alpha')^{-1/2} = g\bar M_{P}$) yield supergravity as a
low-energy effective field theory with infinite towers of
higher-dimension operators with coefficient functions proportional to inverse
powers of $M_{str}$ and powers of the compactification scale (the latter may
be non-explicit in four-dimension string formulations).  One must
therefore take account of higher-dimension operators when analyzing terms
which contribute to fermion mass matrices.  This was, indeed, already realized
long ago \cite{eg}, although detailed studies have only been carried out
recently. At first, one
might consider these higher-dimension operators to be an unfortunate, if
inevitable, complication in the theory.  However, they may well play a very
important role in the area of fermion masses.  Specifically, via vacuum
expectation values (vev's) of the scalar components of certain chiral
superfields, which we shall denote generically as $v$, these
higher-dimension operators can yield contributions to effective
dimension-4 Yukawa interactions which are suppressed by powers of the ratio
$\epsilon \sim v/M_{str}$.  This idea has already been used for a possible
explanation of properties (i) and (iv) \cite{ir}; here we shall extend
this work with new solutions of the consistency conditions and take a
step further, to use the small ratio $\epsilon$ to explain (ii) and (iii).
A preliminary report of some of our results was given in Ref. \cite{fnal}.
Related results in a somewhat different direction (having $m_b$ and $m_\tau$
arise from dimension-4 operators, as in Ref. \cite{ir}) were presented in
Ref. \cite{pierre}.

   We shall work within the context of a supergravity theory which
reduces at low energies to the minimal supersymmetric standard model (MSSM).
We consider a theory where there is a flavor and generational
symmetry group $G_F$ which restricts the forms of the terms in the
action.  In particular, this symmetry forbids certain cubic superfield
couplings which give rise to Yukawa interactions.  Since this happens at a
scale not too far from that of quantum gravity, and since, in general, global
symmetries are broken by quantum gravity \cite{qgbk} even at the semi-classical
level, one is motivated to make $G_F$ a gauge symmetry.  One then faces two
questions: (a) is there a natural origin for $G_F$ in the presumed underlying
string theory? and (b) is there a natural way to explain why the scale of the
breaking of $G_F$ is such that the ratio $v/M_{str}$ has the value that it
must
to fit the observed forms of fermion mass matrices?  A possible affirmative
answer to both of these questions is provided by a gauged symmetry $G_F=U(1)_A$
which is, at the field theory level, apparently anomalous, but whose anomaly is
cancelled by a Green-Schwarz mechanism \cite{gs}. Such U(1)$_A$ gauge
symmetries are known to arise is various string models and, moreover, they
are broken at a calculable scale $v$ given by
$v^2 \simeq M_{str}^2/(192 \pi^2)$ \cite{dsw}, so that
$\epsilon \sim (8 \pi \sqrt{3})^{-1} = 0.023$, a value which is in the
right general range to explain fermion mass hierarchies \cite{ir}.
Of course, such a U(1)$_A$ symmetry does not mix up-type and down-type quark
superfields, or mix these with lepton superfields and is thus quite different
from flavor and generational symmetries which comprise extensions of
grand unified groups.

   We shall denote the (left-handed) SM matter chiral superfields as
$Q_i$, $u^c_i$, $d^c_i$, $L_i$, and $e^c_i$, where $i=1,2,3$ is the
generation, with $\{u_i \} = \{u,c,t \}$, $\{d_i \} = \{d,s,b \}$, etc.
Under the flavor- and generation-dependent U(1)$_A$, these carry the
charges $q_{_{Q_i}}$, $q_{u^c_i}$, $q_{d^c_i}$, $q_{_{L_i}}$, and
$q_{e^c_i}$.  The $Y=1,-1$ Higgs chiral superfields are denoted $H_1$ and
$H_2$ and have U(1)$_A$ charges $q_{_{H_1}}$ and $q_{_{H_2}}$.
  We shall also assume that the theory is invariant under the
usual $R$ parity.  The cubic superfield terms in the globally supersymmetric
superpotential are then given by
\beq
W_{cubic} = Q_i Y^{(u)}_{ij} u^c_j H_2 +  Q_i Y^{(d)}_{ij} d^c_j\ H_1 +
L_i Y^{(e)}_{ij} e^c_j H_1
\label{wcubic}
\eeq
We shall assume that the gauge symmetry at energies $\gtwid v$ is $G=G_{SM}
\times$ U(1)$_A$, where $G_{SM}$ is the SM gauge group SU(3) $\times$ SU(2)
$\times$ U(1)$_Y$.  The theory will also contain certain chiral superfields
which are SM-singlets but transform under the U(1)$_A$.  A discussion of
 constraints on these to avoid destabilization of the gauge hierarchy is given
in Ref. \cite{vj}.

   To proceed, we recall how the cancellation of an apparent field-theoretic
anomaly in a
gauged U(1) symmetry works when this U(1) is, in fact, non-anomalous in the
full string theory (Green-Schwarz mechanism) \cite{gs} and the related
(Dine-Seiberg-Witten, DSW) mechanism whereby the U(1) is
broken \cite{dsw}. Here, we
shall denote chiral superfields generically by $\Phi$.
Standard $d=4$ supergravity is described by two functions. The first is the
generalized K\"ahler  potential
\beq
G(\Phi,\bar\Phi)=K(\Phi,\Phi)+\ln W(\Phi)+\ln \bar W(\bar\Phi)
\label{geq}
\eeq
is a hermitian function of the K\"ahler potential $K$ and the
superpotential $W$, the latter being a holomorphic function of the
chiral superfields. The second is the gauge kinetic normalization
(matrix) function $f_{ab}(\Phi)$, where $a,b=1,..,N_G$, the number of gauge
bosons.  As indicated, $f_{ab}$ is a holomorphic function of the chiral
superfields.  We shall use $G$, $K$, and $W$ to denote both superfield
quantities and also their scalar components; however we shall distinguish
between chiral superfields and their scalar components by type case
(e.g. $\phi=\Phi|$, $m=M|$).

   The bosonic part of the standard supergravity lagrangian is \cite{sugra}
\beq
{1\over\sqrt{g}}{\cal L}_B = G_{i\bar{j}}D_\mu \phi^i g^{\mu\nu} D_\nu
    \phi^{\bar{j}}+{1\over2}R-{1\over4}{\rm Re}f_{ab}F_{\mu\nu}^aF^{b\mu\nu}
    -{1\over4}{\rm Im}f_{ab} F_{\mu\nu}^a\tilde{F}^{b\mu\nu}-V,
\label{lsugrabosonic}
\eeq
where the auxiliary field contributions to the potential are given by
$V = \hat{V}+{\cal D}$, with
\beq
\hat{V} =  e^G ( G_i G_{\bar{j}}G^{i\bar{j}}-3 )
\label{vhat}
\eeq
and
\beq
{\cal D} = {\rm Re}(f_{ab}^{-1})D^a D^b \ ,
\label{d}
\eeq
where $D_a$ is the D-type auxiliary field associated with generators $T_a$ of
the gauge group, normalized
according to $Tr[T_a,T_b]=(1/2)\delta_{ab}$.
Discussions of one-loop corrections are in Ref. \cite{gj} and references
therein.
The scalar fields and their conjugates are denoted as above by $\phi^i$ and
$\phi^{\bar{i}}=\overline{\phi^i}$.  The K\"{a}hler metric
$G_{i\bar{j}}=K_{i\bar{j}}=\partial_{\phi^i}\partial_{\phi^{\bar{j}}}$ has
inverse $G^{i\bar{j}}$. Finally, the derivatives $D_\mu$, which are gauge
and general coordinate invariant, are normalized so that $D_\mu=\partial_\mu
+\sum_a iA^a_\mu T_a$.

The function $f_{ab}$ determines the gauge coupling constants of the theory and
also plays an important role in the cancellation of the field
theoretic anomaly due to the $F \tilde F$ term in the Lagrangian.
   In string models, at tree level, $f_{ab}$ is given by
\beq
 f_{ab}=k_a s \delta_{ab},
\label{fab}
\eeq
where the levels $k_a$ of the Kac-Moody algebras on the worldsheet depend on
the gauge factor group $G_a$ \cite{kmlevel}.  We denote $k_i$, $i=1,2,3$ and
$k_A$ respectively as the Kac-Moody levels corresponding to the factor groups
U(1)$_Y$, SU(2), and SU(3) of $G_{SM}$ and the flavor symmetry U(1)$_A$.
The dilaton vev determines the gauge couplings, according to
$1/g_a^2=k_a<{\rm Re}(s)>$.  At one- and higher-loop levels, $f_{ab}$ acquires
a dependence also on the moduli fields \cite{kl,ant,il}
 from chirally anomalous triangle diagrams in the effective field theory as
well as string threshold corrections generically needed to cancel these
anomalous terms, as is required by the underlying string theory.
However, there is a
large class of models in which the
field theoretic chiral anomalies can be cancelled by a ``universal''
Green-Schwarz mechanism involving the dilaton, similar to the gauge $U(1)$
anomaly cancellation, in which the one-loop corrected $f_{ab}$ is such that
$Re(f_{ab})$ (which is what controls the unification conditions for the gauge
couplings) has the form $Re(f_{ab}) = k_a\delta_{ab}(Re(s) + f(T,\bar T))$,
where $T$ denotes the moduli \cite{kl,ant,il}.
We shall restrict our attention to this class of models.  (There are
also small logarithmic corrections from the running of the gauge couplings
between $M_{str}$ and $M_X$, the scale of gauge coupling unification;
as in \cite{ir}, we shall neglect these here
since $M_X$ is not $<< M_{str}$.)

 If the effective theory contains no gauge anomalies, then
(\ref{lsugrabosonic}) is gauge-invariant, with the dilaton being a gauge
singlet. In the case of interest here, however, where $G$ contains a U(1)$_A$
factor which has field theoretic anomalies, then one-loop corrections from
light ($<\bar M_{Pl}$) fermions give an anomalous correction ${\cal L}_{anom}$
to (\ref{lsugrabosonic}).  Under a U(1)$_A$ transformation,
${\cal L}_{anom}$ transforms by
\beq
\delta {\cal L}_{anom}\propto c_A F_A^{\mu\nu}\tilde{F}^A_{\mu\nu}
     +\sum_a c_a F_a^{\mu\nu}\tilde{F}^a_{\mu\nu}+...,
\label{delta_l}
\eeq
where
\beq
c_A = (1/3){\rm Tr}(T_A T_A T_A)
\label{caaa}
\eeq
\beq
c_a = {\rm Tr}(T_A T_a T_a),
\label{ca}
\eeq
and $a$ runs over the gauge group factors which have no field
theoretic anomalies (in the present case, the factors in $G_{SM}$).
The ellipses in (5) indicate additional terms which we shall discuss below.

   The theory is invariant under constant rescalings $A^\mu_a \rightarrow
\alpha A^\mu_a$, which amounts to a simultaneous redefinition of $k_a$ and
$T_a$: $k_a\rightarrow \alpha^2 k_a$ and $T_a\rightarrow \alpha T_a$. (Our
normalization of the generators of $G_{SM}$ is the standard one.) For the
$U(1)_A$ group the above rescaling could be used to set $k_A$ to 1 at the
expense of modifying the $U(1)_A$ charges. We shall instead use this rescaling
to fix the $U(1)_A$ charge of one of the SM-singlet particles.

  Due to the coupling of Im($s$) to $F\tilde{F}$ in (\ref{lsugrabosonic}),
 this variation of ${\cal L}_{anom}$ can be cancelled by assigning a
$U(1)_A$ gauge variation to Im($s$) \cite{gs,dsw}, but only if
\beq
c_A : c_a :: k_A : k_a \quad {\rm for}\:\:{\rm all}\;\; a.
\label{krelation}
\eeq
Since $s$ is no longer a gauge singlet, its kinetic terms must be made gauge
invariant.  At the superfield level, this means that the K\"ahler function
$K$ must include a coupling between the chiral superfield $S$ and $V_A$, the
real superfield that contains the $U(1)_A$ gauge multiplet.  Specifically,
the tree level K\"ahler function for $S$ is modified as \cite{dsw}
\beqs
-\ln(S+\bar{S}) &\rightarrow & -\ln(S+\bar{S}+cV_A) \nonumber \\
   &=& -\ln(S+\bar{S}) - {cV_A\over S+\bar{S}} + {c^2 V_A^2\over
    2(S+\bar{S})^2 } + ...
\label{kalmod}
\eeqs
where $c$ is a constant determined by gauge invariance and is
related to the coefficients (\ref{ca}). The second term in the last line of
(\ref{kalmod}) produces a term in ${\cal L}_B$ which is linear in the
auxiliary $D$ field of $V_A$.  Once $Re(s)$ acquires a nonzero vev, this causes
some of the fields which are charged under the U(1)$_A$ to acquire vev's,
thereby spontaneously breaking U(1)$_A$.

In a given string model, (\ref{krelation}) is automatically satisfied, due to
the absence of gauge anomalies in the underlying theory. The absence of
anomalous terms that mix the different gauge group factors requires that
\beq
{\rm Tr}(T_A T_A T_a) =0 \quad {\rm for}\;\;{\rm all}\;\; a.
\label{mixedanom}
\eeq
This ensures that $\delta {\cal L}_{anom}$ does not contain any
$F\tilde{F}$ terms that cannot be cancelled by a gauge variation of $s$.
Furthermore, mixed gauge and gravitational anomalies must be absent.
In addition to terms explicitly displayed in (\ref{kalmod}), the U(1)$_A$
gauge variation of ${\cal L}_{anom}$ contains a term proportional to
Tr$(T_A) R^{\mu\nu}\tilde{R}_{\mu\nu}$. This term can be cancelled because
the effective Lagrangian contains a nonstandard (higher derivative)
coupling $k_G{\rm Im}(s) R^{\mu\nu}\tilde{R}_{\mu\nu}$ (which defines $k_G$).
The cancellation requires a
relation between the coefficients $c_a,c_A$  and Tr$(T_A)$ because the same
$U(1)_A$ transformation of $s$ must cancel all anomalies. This last relation
is difficult to use phenomenologically because (i) Tr$(T_A)$ depends on SM
singlets about which we have very little experimental information, and (ii)
because it depends on $k_A$ and $k_G$ which are not measured at low
energies.

We proceed to investigate such models to obtain phenomenologically
acceptable fermion mass matrices.
Within the above theoretical framework, we shall make the following specific
assumptions:

\begin{enumerate}

\item The low energy theory near the electroweak scale is the MSSM with
phenomenologically viable soft supersymmetry breaking and supersymmetric
mass terms. This assumption precludes SM nonsinglets which carry $U(1)_A$
charge from getting large vev's due to the $U(1)_A$ $D$--term.

\item

Supersymmetry is spontaneously broken in a hidden sector with \newline
$m_{3/2}^2= \ <e^G> \ \sim m_W^2$ and $<V> \ \sim 0$; i.e., $O(M_P^4)$
and $O(M_P^2 m_{3/2}^2)$ contributions to $<V>$ cancel.
This ensures that the soft breaking terms are naturally of $O(m_W)$. In
addition, this assumption means that the dominant contributions to the
effective Yukawa couplings are only from superpotential terms, as
in the MSSM, and not from $K$, which can also potentially
contribute to the effective Yukawa couplings in
a supergravity model \cite{js2}.

\item
The SM gauge couplings unify in the canonical way, i.e., $g_3^{-2} :
g_2^{-2} : g_Y^{-2} = 1:1:5/3$, or equivalently, $k_3:k_2:k_1=1:1:5/3$.

\item
SM singlet fields $\chi$ which get a vev with $<D_A> \ =0$ carry
$U(1)_A$ charges with the same sign.  In the class of models that we consider,
$<\chi> \ \sim 10^{16}$ GeV, which is $ >> m_W$. This ensures that $<W>$
(and hence
$<e^G>$) can be kept small without resorting to accidental near-cancellation
between two large vev's in $<W>$ or introducing extra symmetry. For example,
if the $U(1)_A$ charge of $\chi$ is $1$ and the $U(1)_A$ charge of another
SM singlet $\chi'$ is $-b$, $b=2,3,...$, then gauge invariance allows $W$ to
contain terms of the form
$\chi^b\chi'/\bar M_P^{(b-2)}$. If both $\chi$ and $\chi'$ get vev's $\sim
\bar M_P/100$ then the contribution of this term to $<W>$ is phenomenologically
too large, unless $b$ is sufficiently large.

\item
To reduce the number of parameters, it is convenient to assume that the
effective Yukawa matrices are symmetric at $M_X$.

\end{enumerate}

The hierarchical structure of the low-energy fermion masses can be
produced by Yukawa couplings which are either hierarchical or democratic.
In the latter case, all entries of a Yukawa matrix are the same to leading
order, while subleading corrections to these differ so that the eigenvalues
of the matrix have the desired hierarchical structure. A $U(1)_A$ symmetry can
potentially explain hierarchical Yukawa matrices but not (by itself) democratic
matrices. In contrast, a hierarchical structure can be produced by assigning
$U(1)_A$ charges to the operators in (\ref{wcubic}) in a manner such that
for different $i,j$, these couple to different requisite powers of
certain SM singlet chiral superfields.

   We next determine which U(1)$_A$ charge assignments for the fields satisfy
the various constraints and lead to experimentally viable fermion Yukawa
matrices.  There are $N_q=5N_G+2=17$ \ U(1)$_A$ charges in the MSSM, where
$N_G=3$ is the number of matter generations. In addition, there are $N_s$ \
U(1)$_A$ charges for SM-singlet chiral superfields.
The conditions (\ref{krelation}) and (\ref{mixedanom}), together
with the assumption of symmetric mass matrices, reduce these 17 parameters to
8.  To show this, we first list the U(1)$_A$ charges of the SM matter fields
as in Table \ref{table1}, where $\bar q_{f} = (1/3)\sum_{i=1}^3 q_{f_i}$ is the
generational average charge for the $f$ chiral superfield.
For the Higgs chiral superfields, we put $q(H_i) = q_{_{H_i}}$, $i=1,2$.
\begin{table}
\begin{center}
\begin{tabular}{|c|c|c|c|c|c|} \hline\hline
 $f$ & $Q$ & $u^c$ & $d^c$ & $L$ & $e^c$ \\ \hline\hline
$q_{_{f_1}}$ & $\bar q_{_{Q}} +\alpha_1$ & $\bar q_{u^c} +\beta_1$
& $\bar q_{d^c} +\gamma_1$ & $\bar q_{_{L}} + a_1$ & $\bar q_{e^c} + b_1$ \\
$q_{{f_2}}$ & $\bar q_{_{Q}} +\alpha_2$ & $\bar q_{u^c} +\beta_2$
       & $\bar q_{d^c} +\gamma_2$ & $\bar q_{_{L}} + a_2$
       & $\bar q_{e^c} + b_2$ \\
$q_{{f_3}}$ & $\bar q_{_{Q}} -(\alpha_1 + \alpha_2)$
       & $\bar q_{u^c} -(\beta_1 + \beta_2)$
       & $\bar q_{d^c} - (\gamma_1 + \gamma_2)$
       & $\bar q_{_{L}} -(a_1+a_2)$
       & $\bar q_{e^c} -(b_1+b_2)$ \\ \hline
\end{tabular}
\end{center}
\caption{U(1)$_A$ charges of SM matter chiral superfields $f_i$. }
\label{table1}
\end{table}
To ensure symmetric Yukawa matrices, we require that the U(1)$_A$ charges of
the chiral superfield bilinears satisfy
\beq
q(Q_i u^c_j) = q(Q_j u^c_i)
\label{qusym}
\eeq
\beq
q(Q_id^c_j) = q(Q_j d^c_i)
\label{qdsym}
\eeq
\beq
q(L_i e^c_j)=q(L_j e^c_i)
\label{lsym}
\eeq
In all, there are six independent contraints in these equations.
The solutions are
\beq
\alpha_i = \beta_i = \gamma_i \ , \qquad a_i = b_i \ , \ for \ \ i=1,2
\label{symmsol}
\eeq
Because the anomalies $c_i$ in eq. (\ref{ca}) are linear in the U(1)$_A$
charges, it follows that, regarding the contributions from the matter fields,
the $c_i$ only depend on the generational average U(1)$_A$ charges.
These anomalies are
\beq
U(1)_Y^2U(1)_A \ : \ (1/4) \Bigl [ N_G \Bigl \{ (2/3)\bar q_{_{Q}} +
(16/3)\bar q_{u^c} + (4/3)\bar q_{d^c} + 2\bar q_{_{L}} + 4\bar q_{e^c}
\Bigr \} +2q_{_{H_1}}+2q_{_{H_2}} \Bigr ]
\label{c1}
\eeq
\beq
SU(2)^2U(1)_A \ : \ c_2 = (1/2)\Bigl [ N_G(3\bar q_{_{Q}} + \bar q_{_{L}}) +
q_{_{H_1}} + q_{_{H_2}} \Bigr ]
\label{c2}
\eeq
\beq
SU(3)^2U(1)_A \ : \ c_3 = (N_G/2)(2\bar q_{_{Q}} + \bar q_{u^c} + \bar q_{d^c})
\label{c3}
\eeq
where we show the general $N_G$ dependence but take $N_G=3$ here. The
anomaly conditions (\ref{krelation}) yield two linearly independent
constraints on the 7 parameters $\bar q_{_{Q}}$, $\bar q_{u^c}$,
$\bar q_{d^c}$, $\bar q_{_{L}}$, $\bar q_{e^c}$, $q_{_{H_1}}$, and
$q_{_{H_2}}$, e.g., $c_2=c_3$ and $c_1 = (5/3)c_2$.  These can be solved in
terms of the 5 quantities $x$, $y$, $z$, $v$, and $w$ according to
\beq
\bar q_{_{Q}}=v+x,\quad \bar q_{u^c}=2v+x,\quad \bar q_{d^c}=w+y,
\quad \bar q_{_{L}}=y, \quad \bar q_{e^c}=x, \quad q_{_{H_1}}=3(v+w)+z,
\quad q_{_{H_2}}=-z
\label{qsols}
\eeq
The U(1)$_Y$U(1)$_A^2$ anomaly (\ref{cyaa}) is quadratic in the
U(1)$_A$ charges, and hence, in general, cannot be written just as a function
of the generational averages of the matter field charges. However, given
(\ref{qusym})-(\ref{lsym}), this anomaly also depends only on these averages;
requiring that it vanish gives
\beq
U(1)_Y U(1)_A^2 \ : \ 0 = c_{_{YAA}} = N_G(\bar q_{_{Q}}^2 - 2\bar q_{u^c}^2
+ \bar q_{d^c}^2 - \bar q_{_{L}}^2 + \bar q_{e^c}^2) + q_{_{H_2}}^2 -
q_{_{H_1}}^2
\label{cyaa}
\eeq
In terms of the 5 quantities in eq. (\ref{qsols}), eq. (\ref{cyaa}) is
\beq
0 = 2(w^2-v^2) + 3v(w-x) + vz + w(y+z)
\label{cyaasol}
\eeq
We find the following three families of solutions to (\ref{cyaasol}), which
are thus solutions to the total set of anomaly constraints (these solutions are
independent of $N_s$):
\beq
\begin{array}{ccccccc}
\bar q_Q & \bar q_{u^c} & \bar q_{d^c} & \bar q_L & \bar q_{e^c}
& q_{_{H_2}} & q_{_{H_1}} \\
  x & x & y & y & x & z & -z \\
  x & x & {1\over 2}(y-z) & y & x & -{1 \over 2}(3y+z) & -z \\
  \; x+v \; & \; x+2v \; & \; y+w \; & \; y \; & \; x \; &
   \; 3(v+w)+z \; & \; -z \;
\end{array}
\label{solfamilies}
\eeq
The first two correspond to $v=0$ and describe two distinct 3-parameter
families of solutions.  The last exists for $v \neq 0$ and describes a
4-parameter family of solutions with $x$ solved for in (\ref{cyaasol}).
The first solution in (\ref{solfamilies}) was already given in
Ref. \cite{ir}; the other two were not mentioned there and are a new
result in the present work.
The four parameters $\alpha_1,\alpha_2,a_1,a_2$,  together with the
unknown parameters in (\ref{solfamilies}) yield all allowed U(1)$_A$ SM
charges consistent with our assumptions. We find that the constraints are very
restrictive, as will be seen.

   In order to account for the important feature (ii) that $m_t$ is comparable
to the EWSB scale $v_{_{EW}}$, one chooses the source of $m_t$ to be a
renormalizable, dimension-4 Yukawa coupling, as in the SM. This requires the
U(1)$_A$ charge of $Q_3u^c_3H_2$ to be zero, i.e.,
\beq
\bar q_{_{Q}}+\bar q_{u^c}+q_{_{H_2}}=2(\alpha_1+\alpha_2),
\label{yu33cond}
\eeq
Let us denote the U(1)$_A$ charge of $(Q_3 d^c_3 H_1)$ as $\theta$. Then
\beq
\bar q_{_{Q}}+ \bar q_{d^c}+q_{_{H_1}}=2(\alpha_1+\alpha_2)+\theta.
\label{yd33cond}
\eeq
Under this assumption, the $U(1)_A$ charges of $(Q_i u^c_j H_2)$
are given by the matrix
\beq
 \( \begin{array}{ccc}
        \; 4\alpha_1+2\alpha_2 \; & \; 3\alpha_1+3\alpha_2\; &
        2\alpha_1+\alpha_2 \\
      3\alpha_1+3\alpha_2 & 4\alpha_2+2\alpha_1 & 2\alpha_2 +\alpha_1 \\
     2\alpha_1+\alpha_2 & 2\alpha_2+\alpha_1 & 0 \end{array}
  \)
\label{upqmatrix}
\eeq
while
\beq
q(Q_i d^c_j H_1) = q(Q_i u^c_j H_2)+\theta.
\label{dnqmatrix}
\eeq
If one took $\theta=0$, then the $b$ quark mass would arise from a
renormalizable cubic superfield operator, and one would not have any
fundamental explanation of property (iii).  The origin of the large mass
ratio $m_t/m_b$, rather than being explained naturally, would have to be
pushed into a similarly large value of $\tan \beta = v_2/v_1$, viz.,
$\tan \beta \sim m_t/m_b$.
Instead, we take $\theta \ne 0$, implementing our explanation of properties
(ii) and (iii), since then $m_b$ arises from higher-dimension operators, and
$m_b << m_t$ follows naturally.

     We proceed to construct an explicit model for fermion mass matrices
embodying these ideas.  We assume that there are $N_s=2$ chiral
superfields which are SM singlets (and $U(1)_A$ nonsinglets),
$\chi$ and $\chi'$ with unequal $U(1)_A$ charges $q_{\chi}$ and $q_{\chi'}$.
For convenience, let us normalize $q_{\chi}=1$ (this can be done by
rescaling the U(1)$_A$ coupling) and define $q_{\chi'} \equiv \theta'$. The DSW
breaking of the U(1)$_A$ yields values for $<\chi>/\bar M_P \ \sim \
<\chi'>/\bar M_P$ (denoted $\epsilon$ above) which are $ \sim O(\lambda^2)$,
where $\lambda=|V_{us}|
\simeq 0.22$ is a measure of the hierarchical structure of the CKM matrix.
First, consider the up- and down-quark masses. We have
\beq
W\ni {\cal Y}^u_{ij}(\chi,\chi')H_2Q_iu^c_j
    +{\cal Y}^d_{ij}(\chi,\chi')H_1 Q_i d^c_j,
\label{wform}
\eeq
so that $Y^u_{ij}=<{\cal Y}^u_{ij}>$ and $Y^d_{ij}=<{\cal Y}^d_{ij}>$.
Motivated by string theory considerations, we shall consider only functions
${\cal Y}$ which do not contain fractional or negative powers of $\chi$ and
$\chi'$.
In order to obtain a viable form of the quark mass matrices, we
require $Y^u_{22}\sim\lambda^4$, i.e. (${\cal Y}^u_{22}\sim \chi^2,
\chi'^2$ or $\chi'\chi$) or ($Y^u_{23}\sim\lambda^2$, i.e.
${\cal Y}^u_{23}\sim\chi$ or $\chi'$).

We consider three cases: \\
(1) ${\cal Y}^u_{22}\sim\chi\chi'$. U(1)$_A$ charge conservation (CC) then
implies
${\cal Y}^u_{23}={\cal Y}^u_{32}\neq\chi,\chi'$. \\
(2) ${\cal Y}^u_{23}\sim\chi$. Then CC $\Rightarrow$
${\cal Y}^u_{22}\sim\chi^2$.\\
(3) ${\cal Y}^u_{23}\sim\chi'$. Then CC $\Rightarrow$
${\cal Y}^u_{22}\sim\chi'^2$.

We have studied all of these cases and find in case (2) an assignment of
U(1)$_A$ charges which yields effective Yukawa matrices (i.e., the matrices
which enter in effective dimension-4 Yukawa terms, all of which, except for
$Y^u_{33}$, actually arise from higher-dimension operators)
which give an acceptable pattern of fermion masses at the
electroweak scale. At $M_X$, these are close to the simple forms
\beq
Y^{(u)} \propto  \left (\begin{array}{ccc}
                  0 & \lambda^6 & 0 \\
                  \lambda^6 & 0 & \lambda^2 \\
                  0 & \lambda^2 & 1 \end{array}   \right  ), \qquad\qquad
Y^{(d)} \propto   \left ( \begin{array}{ccc}
                  0 &  \lambda^4 &  0 \\
                  \lambda^4 & \lambda^3 & \lambda^3 \\
                  0 & \lambda^3 & 1 \end{array}  \right ) ,
\label{yd2}
\eeq
where the actual entries in the positions given by zeros in (\ref{yd2})
need not be, and are not in general, exactly zero; indeed, one may have
$Y^{(u)}_{22}\sim\lambda^4$,
$Y^{(u)}_{11} \ltwid O(\lambda^8)$, $Y^{(u)}_{13} \ltwid O(\lambda^4)$,
and so on (e.g. \cite{ir}).  The solution we give below satisfies
these bounds. In writing such forms, it is understood that the coefficients
$a_{ij}$ multiplying a given power of $\lambda$ may differ from unity,
but not by as much as a positive or negative integer power of $\lambda$.
The pattern (\ref{yd2}) is known to be experimentally viable \cite{rrr,ks}, and
our U(1)$_A$ charge assignments constitute a new way of obtaining this pattern.

Consider two cases \\
(2a) $\theta=-1$. Then, as for case (1a), ${\cal Y}^d_{ij}=\chi{\cal Y}^u_{ij}
  +\tilde{{\cal Y}}^d_{ij}$. Since $Y^u_{22}\sim\lambda^4,$ $Y^u_{23}=
Y^u_{32}\sim\lambda^2$, $Y^u_{33}\sim 1$ and $Y^d_{23}=Y^d_{32}\sim
\lambda^4+<\tilde{{\cal Y}}^d_{23}>$, we must fit (\ref{yd2}).
We have $Y^d_{22}\sim\lambda^6+<\tilde{{\cal Y}}^d_{22}>$
and we need $Y^d_{22}\sim \lambda^4$, so we require $\tilde{{\cal Y}}^d_{22}
\sim \chi^b\chi'^{(2-b)}$ for $b-0,1,2$. Since $4\alpha_2+2\alpha_1=-2$,
CC $\Rightarrow$
\beq
b+\theta'(2-b)-3=0.
\label{btheta}
\eeq
This cannot be satisfied for $b=2$, while for $b=1$, $\theta'=2$ which
would mean ${\cal Y}^u_{22}\sim \chi',\chi^2$ or $Y^u_{22}\sim \lambda^2+...$
which is too big. Hence, the only acceptable solution is
\beq
b=0,\quad \theta'=3/2.
\label{fsol}
\eeq
We also require $Y^u_{12}\sim\lambda^6$, i.e. ${\cal Y}^u_{12}\sim
\chi^a\chi'^{(3-a)}$ for $a=0,1,2,3$, and $Y^d_{12}\sim\lambda^6$, i.e.
$\tilde{{\cal Y}}^d_{12}\sim\chi^c\chi'^{(3-c)}$ for $c=0,1,2,3$. This
leads to the CC equations
\beq
a+(3/2)(3-a)-3\alpha_2-3=0,
\label{cc1}
\eeq
\beq
c+(3/2)(3-c)-3\alpha_2-4=0,
\label{cc2}
\eeq
which immediately implies $a=c+2$. Since $a$ cannot be bigger than $3$ we
need only consider the cases $c=0$ or $c=1$. Thus we have the solutions
\beq
b=0,\ c=0, \ a=2, \ \alpha_2=1/6, \ \alpha_1=-4/3, \ \theta'=3/2,
\label{fullsol1}
\eeq
and
\beq b=0, \ c=1, \ a=3,\alpha_2=0, \ \alpha_1=-1, \ \theta'=3/2.
\label{fullsol2}
\eeq
The last case allows ${\cal Y}^u_{12}\sim\chi'^2$, i.e. $Y^u_{12}\sim
\lambda^4$ which is too big. Therefore the only viable solution
corresponds to the first set, eq. (\ref{fullsol1}).
 For this solution, we find
\beq
Y^u \sim  \left (\begin{array}{ccc}
                  \lambda^8 & \lambda^6 & \lambda^4 \\
                  \lambda^6 & \lambda^4 & \lambda^2 \\
                  \lambda^4 & \lambda^2 & 1 \end{array}   \right  ),
\qquad\qquad
Y^d \sim  \lambda^2  \left ( \begin{array}{ccc}
                  \lambda^6 &  \lambda^4 &  \lambda^4 \\
                  \lambda^4 & \lambda^2 & \lambda^2 \\
                  \lambda^4 & \lambda^2 & 1 \end{array}  \right );
\label{ydd2}
\eeq
where, for example, the coefficients of $Y^u_{33}$, $Y^d_{23}$,
$Y^d_{32}$ and $Y^d_{22}$ terms could be small enough
to satisfy experimental bounds, and so forth for other entries.

   Our explanation for property (iii) also requires that the entire
contribution to the lepton matrices arises from higher-dimension operators.  A
charge assignment which satisfies this requirement is
\beq
  \bar q_{_{L}}+\bar q_{e^c}+q_{_{H_1}} = 2(a_1+a_2)-1,
\label{ye33cond}
\eeq
\beq
  a_1=\alpha_1=-4/3,\quad a_2=\alpha_2=1/6.
\label{cass}
\eeq
These equations imply that the charge assignments for the lepton sector
are as for the down-quark sector, i.e.
$Y^e \sim Y^d$ at $M_X$. This allows $m_{\tau}\approx m_b$
and $m_{\tau}m_{\mu}m_e\approx m_d m_s m_b$ at $M_X$ and, given
the above-mentioned freedom in the coefficients of the powers of $\lambda$,
this can produce a viable model for lepton masses.\footnote{Since a factor of 3
falls within the accepted range $\lambda < a_{ij} < \lambda^{-1}$, the model of
Ref. \cite{ksmodel} is an illustration of how one can fit experiment with $Y^e
\sim Y^d$.}

Given the charge assignments (\ref{cass}), the three linear equations
(\ref{yu33cond}),(\ref{yd33cond}) and (\ref{ye33cond}) can be used to
further restrict  the solutions to the anomaly constraints (\ref{solfamilies}).
For example, for the first 3-parameter family of solutions we require
$x=-(1/6)(3z+7)$ and $y=(1/6)(9z-13)$.

  To pursue this line of research further, the next step is to investigate
how the U(1)$_A$ charge assignments that we have made can be derived from a
deeper theory (presumably the underlying string theory).  Another topic for
study, but one with much weaker constraints, is that of neutrino masses and
mixing. Further details will be given in Ref. \cite{js2}.
This research was partially supported by the NSF Grant PHY-93-09888.

\vfill
\eject

\end{document}